# ON THE ELECTRICAL AND MAGNETIC PROPERTIES OF SOME INDIAN SPICES

Samson.K.Baby and Girish,T.E
Department of Physics,University College,
Trivandrum 695 034, Kerala
Email: tegirish5@yahoo.co.in

## ABSTRACT

We have made experimental measurements of electrical conductivity ,pH and relative magnetic susceptibility of the aqueous solutions of 24 indian spices. The measured values of electrical conductance of these spices are found to be linearly related to their ash content and bulk calorific values reported in literature. The physiological relevance of the pH and diamagnetic susceptibility of spices when consumed as food or medicine will be also discussed .

## 1. Introduction

Spices are commonly used in India as a part of food preparations and as home medicine for certain ailments. There are several studies related to the biomedical [1,2] and nutritional aspects [3,4] of Indian spices. Biophysical investigations of spices are generally rare. In this paper we report our experimental studies on the electrical ( conductivity and pH) and magnetical ( relative susceptibility) properties of twenty four Indian spices. Electrical conductivity of spices are shown to reflect its mineral content and calorific properties. The physiological relevance of the pH and magnetic susceptibility of spices when consumed as food or medicine will be also discussed.

## 2. Measurements of electrical conductivity, pH and reltive magnetic susceptibility of Indian spices

The list of spices used for our experimental studies are given in Table1. The spice samples are first cleaned and crushed in to be powder form using rotor and pecil arrangement. Aqueous solutions of these spices are prepared using 20g of spice powders and 50 ml of double distilled water. The filtered solutions are used for making electrical and magnetic measurements. The electrical conductivity ( K) and pH are measured using digital conductivity meter and pH meter respectively by periodically checking their calibrations . The relative magnetic susceptibility ( Mr) of the spice solutions are measured using the Quincke's method. given by :

$Mr = Ms/Mw$ where Ms is the dimagnetic susceptibility of the given spice solution and Mw is the diamagnetic susceptibility of the distilled water.

The values of K, pH and Mr of the spice solutions are given in Table 1. For the certain spices like *cardamon* aqueous solution preparations are difficult since they easily get converted in to colloidal form. For these samples magnetic susceptibility measurements are not carried out

## 3. Discussion of results

Apart from use in food preparations spices are included in beverages also. Cardamon and ginger are added to Tea/Coffee for getting good flavour. We have compared our measurements of the electrical conductivity of spice solutions with their ash content and calorific values reported in literature ( available in the website of Spices board of India).

The results are shown in Fig 1 where we could find linear relations between these parameters. High electrical conductivity of spices ( K>10) thus reflect its mineral content or nutrition value.

It is interesting to note that spices which are boiled with drinking water in Kerala households  ( cumin seeds,black pepper,coriander seeds and cardamom) possess relative high electrical conductivity .

Spices can be also classified according to its acidic nature from its pH measurements.Clove is found to be highly acidic ( pH=3.8)  . The pH value of food materials and ayurvedic medicines  are modified by the presence of spices which can help digestion.  For eg. buttermilk is usually prepared  at home by adding coriander/curry leaves , asofoedia  and common salt.

The magnetic  susceptibility of the food is modified by the presence of spices such as garlic which is  found  to posess only half of the diamagnetic susceptibility of  pure water. Garlic is used as a herbal medicine for hypertension. Most of the organic food items are diamagnetic in nature while atmospheric  oxygen we respire is paramagnetic . Human blood is known to contain ferromagnetic substances such as hemoglobin. Thus the "magnetic balance" of the blood circulation  is  likely to be  affected by the magnetic properties of food we consume. Further  the value of the diamagnetic susceptibility  of the human body decides its interaction with external magnetic fields such as the geomagnetic field. Lower the diamagnetic susceptibility better the interaction with external fields. Magneto-biology  is an important area of contemporary scientific research.

**Table1. Electrical conductivity, pH and relative magnetic susceptibility of the aqueous solutions of some Indian spices**

| Name of the spice | Electrical conductivity (mmhos/cm) | pH | Relative Magnetic Susceptibility |
|---|---|---|---|
| 1. Kacholam (*Kaempferia galanga*) | 2.2 | 6 | no data |
| 2. Clove | 10.05 | 3.8 | 0.83 |
| 3. Mace | 4.5 | 4.8 | 1.074 |
| 4. Star Anees | 8.2 | 3.45 | 1.278 |
| 5. Cinnamon | 2.25 | 4.8 | 1.185 |
| 6. Fennel | 14 | 5.75 | 0.849 |
| 7. Nutmeg | 5.3 | 5.75 | 1.538 |
| 8. Bishopweeds | 21.5 | 5.6 | 0.831 |
| 9. Mustard seed | 5.8 | 5.1 | 0.714 |
| 10. Black pepper | 12 | 6.3 | 1.02 |
| 11. Garlic | 3.2 | 5.9 | 0.54 |
| 12. Long pepper | 10 | 6.25 | 0.615 |
| 13. Dill seeds | 12.5 | 6.4 | 1.02 |
| 14. Poppy seeds | 3.7 | 5.5 | 1.2 |
| 15. Dry ginger | 14.5 | 5.9 | 0.913 |
| 16. Curry leaves | 5.1 | 5.75 | 0.78 |
| 17. Mint | 1,3 | 6.7 | 1.41 |
| 18. Corainder leaves | 3.4 | 6.2 | 0.784 |
| 19. Cumin seeds | 13 | 5.8 | 1.06 |
| 20. Green Chilly | 2.3 | 5.6 | 0.94 |
| 21. *Bacopa monneri* | 6 | 5.65 | 0.84 |
| 22. Cardamon | 14.5 | 5.5 | no data |
| 23. Coriander seeds | 11 | 5.25 | -do- |
| 24. Mango Ginger | 5.2 | 5.95 | -do- |

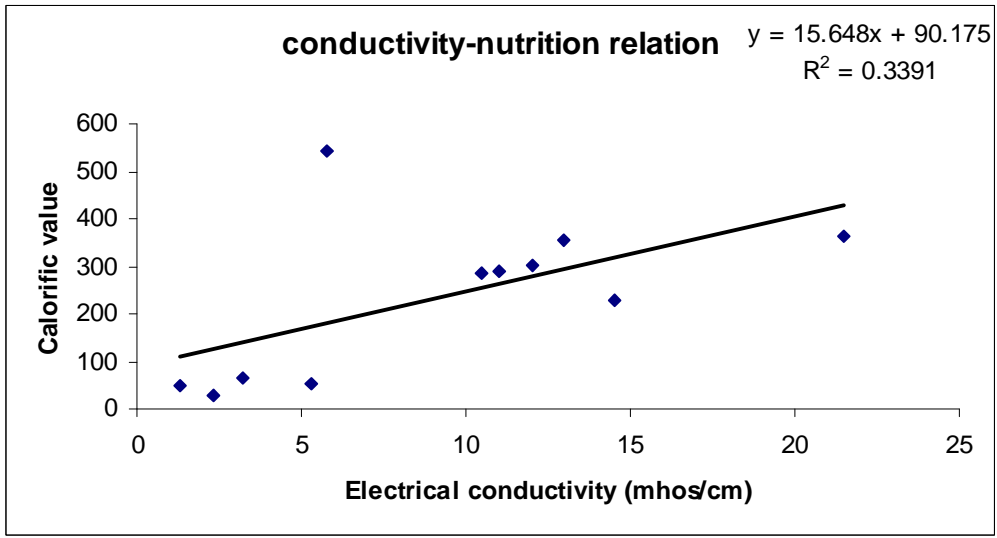

(a)

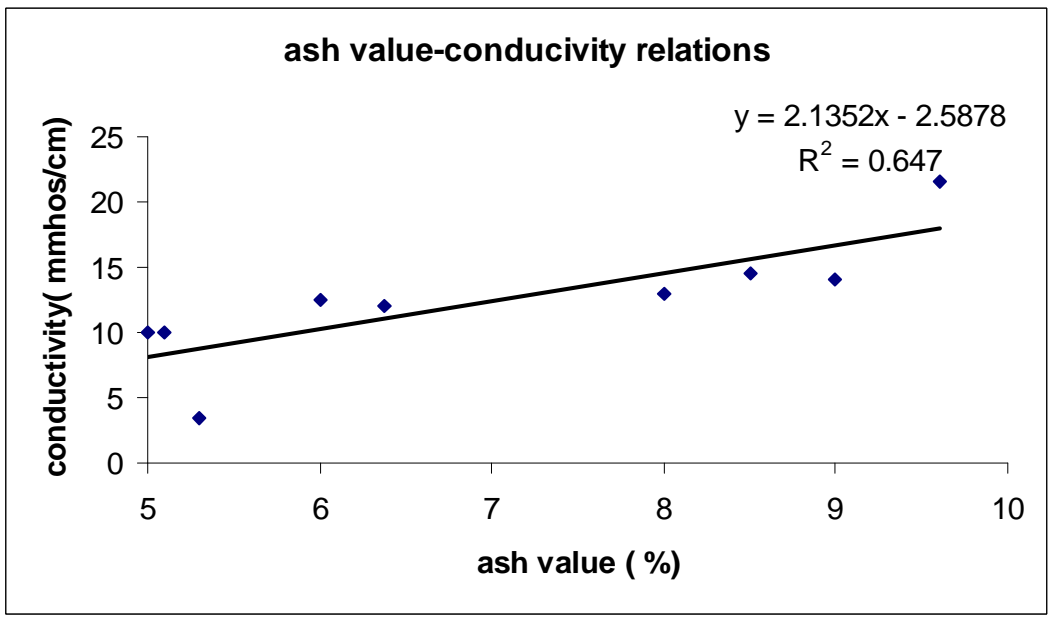

(b)

**Fig 1**. The linear relation between measured electrical conductivity of some Indian spices with (a) ash content  ( b) calorific values  reported in literature.